\documentclass[sn-mathphys]{sn-jnl}
\usepackage{amssymb}

\begin{document}

\title[Article Title]{Indeterminacy in classical cosmology \\with dark matter}

\author{\fnm{Viqar} \sur{Husain}}\email{vhusain@unb.ca}

\author{\fnm{Vladimir} \sur{Tasi\'c}}\email{vtasic@unb.ca}

 \affil{\orgdiv{Department of Mathematics and Statistics}, \orgname{University of New Brunswick}, \orgaddress{\street{}
 \city{Fredericton}, \postcode{E3B 5A3}, \state{NB}, \country{Canada}}}

\abstract{We describe a case of indeterminacy in general relativity for homogeneous and isotropic cosmologies for a class of dark energy fluids. The cosmologies are parametrized by an equation of state variable, with one instance giving the same solution as Norton's mechanical dome. Our example goes beyond previously studied cases in that indeterminacy lies in the evolution of spacetime itself: the onset of the Big Bang is indeterminate. We show further that the indeterminacy is resolved if the dynamics is viewed relationally.}


\maketitle
 
\section{Introduction}

The publication in 2003 of what became known as Norton's Dome \cite{norton1, norton2} generated an interesting debate on determinism in classical dynamics. Related examples had been discussed by other authors: Bhat and Bernstein \cite{bb}
consider a mathematically similar system, Kosyakov \cite{kosyakov} 
includes an analysis of non-uniqueness of solutions of the Cauchy problem
in classical electrodynamics (and independently points to a class of systems mathematically related to Norton's Dome), and Earman \cite{earman} surveys  a broader class of ``fault modes of determinism" in  (non-quantum) 
physics. What distinguishes Norton's approach is the attractiveness of an easily visualized and unexpectedly simple Newtonian (or Newtonian-looking) system. 
 
The dome example \cite{norton1, norton2} involves a point-mass placed at rest at the apex of a dome-shaped surface embedded in $\mathbb{R}^3$ given by the points 
\begin{equation}
 \sigma(r,\phi) = \left(R(r)\cos\phi,R(r)\sin\phi, z_0-\frac{2}{3}r^{3/2} \right)
 \end{equation}
 in cylindrical coordinates $(R,\phi,z)$;  $r$ is the radial arc length at fixed $\phi$ along the dome, measured from its apex  at $z=z_0, R=0$, and $R(r) = \frac{2}{3}\left[1-(1-r)^{3/2} \right]$ (cf. \cite{malament},
 eqn. (4)). The dome is so designed to obtain the Newtonian equation of motion $F=ma$ under gravity
 
\begin{equation}
    \ddot{r}=\sqrt{r} \label{ND}
\end{equation}
for rolling without friction with respect to Newtonian external time $t$. 
With the initial conditions $r(0)=0$ and $\dot{r}(0)=0$, the solution is 
not unique: one solution is $r(t)=0$ for all $t$; another is 
\begin{equation}
r(t) =  \left\{
        \begin{array}{ll}
            0, & t<T  \\
            \frac{1}{144}(t-T)^4, & t \geq T
        \end{array}
    \right.
    \label{d-sol}
\end{equation}
for any $T$. Thus it appears that the particle is at rest at the top of 
the dome up to an arbitrary time $T$, and can then spontaneously undergo 
acceleration for $t\ge T$.  This occurs because the equation (\ref{ND}) 
does not fulfil the Lipschitz condition for uniqueness of solutions.   

Norton \cite{norton2} contends that the dome demonstrates a failure of determinism in Newtonian physics or at least raises questions regarding the widespread views
on determinism in classical mechanics and broadly accepted mathematical idealizations in that context. 
Malament \cite{malament} concludes his extensive  analysis  on a more cautious note:
``we do not have a sufficiently clear idea in the first place what should count as a 
`Newtonian system' (or count as falling within the `domain of application' of Newtonian theory). 
My inclination is to avoid labels here and direct attention, instead, to a rich set of issues that the example raises."

Questions raised by Norton's example have generated a considerable volume literature at the intersection of mathematics, physics and philosophy of science. The literature is too extensive to enumerate in  this note, but includes works  critical of the purported philosophical import of the dome (for instance, Korolev 
\cite{korolev}, Wilson \cite{wilson}, Laraudogoitia \cite{laura2013}), as well as variations that build on Norton's example and connect it 
to other problems in the philosophy of physics (e.g., Lee \cite{lee} strings together an infinite number of domes to construct
a ``simple physical model of the staccato run satisfying all the kinematical and dynamical requirements for a supertask").

The problem has a rich and interesting history. The equation of motion itself, non-uniqueness of its solutions, as well as a slightly more general class of examples, had been considered as early as 1806 by Poisson. Discussions of questions this raises about determinism unfolded at 
various times during the nineteenth century, involving Poisson, Duhamel and others. Fletcher \cite{fletcher1} and
Van Strien \cite{vanstrien} provide historical background. We note
Poisson's later view (as of 1833) regarding the importance
of hypothetical examples and also the difference between the results
of purely mathematical methods and their physical interpretation (\cite{vanstrien}): ``This example, purely hypothetical, suffices to demonstrate the necessity 
to take into account the singular solutions of the differential equations 
of motion, if there were any singular solutions; which does not happen in 
reality [\ldots]".

That mathematical idealizations do not always provide an exact
representation of physical processes is also noted by Arnol'd in  {\it Ordinary Differential Equations} \cite{arnold2}: 
``the form of the differential equation of the 
process, and also the very fact of determinacy [\dots] can be established only by experiment, and consequently only 
with limited accuracy. In what follows we shall not emphasize that circumstance every time, and we shall talk about real 
processes as if they coincided exactly with our idealized mathematical models." 

In response to critiques to the effect that the dome is ``unphysical", or that it cannot be
supported (or refuted) by experiment, Norton \cite{norton2} 
argues that his example is not a statement about the world but about our theories of the  world: 

 \begin{quote}
 The dome is not intended to represent a real physical system. The dome is purely an idealization within 
Newtonian theory. On our best understanding of the world, there can be no such system. For an essential part of the 
setup is to locate the mass {\it exactly} at the apex of the dome and 
{\it exactly} at rest. Quantum mechanics assures us that cannot be done. 
What the dome illustrates is indeterminism within Newtonian theory in an idealized system
that we do not expect to be realized in the world.
\end{quote}

In discussing a variation on the dome, Lee \cite{lee} suggests ``we accept that physical models have varying degrees of metaphysical import depending on what idealizations they employ and that we focus on figuring out the metaphysical implications of such idealizations." It is perhaps in this spirit that systems analogous to Norton's Dome have been constructed for electric charge motion, and for special relativistic and geodesic motion on a fixed curved spacetime (Fletcher \cite{fletcher2}). In each of these cases the resulting equations of  motion are 
equivalent to those for a particle mass on the dome, although the elegant simplicity of Norton's example is difficult to match.

We provide a similar example in cosmology that is a solution of Einstein's field equations.  Unlike the examples 
referenced above, where indterminacy arises from non-uniqueness
of solutions of the equations of motion of an object within a given
spacetime, our example involves  indeterminacy of the evolution
of spacetime itself (more precisely, an extended spacetime).

The cosmological model we present below may  
be viewed at the mathematical 
level as a  variation on Norton's dome, but
one that arises from Friedmann's equations for a range
of the parameter in the matter equation of 
state relating pressure and energy density of a perfect fluid. 
Within the range we discuss to mimic the mathematical effects in Norton's example, values of the parameter correspond to those in a class
of dark energy models.  This 
parameter is therefore not a gauge artifact that can be removed by a 
diffeomorphism of the resulting solutions.   Furthermore, the model's 
solutions may be exhibited as  ``relational" trajectories in phase space 
such  that the indeterminism present with respect to coordinate time is 
removed. 

For these reasons indeterminism in the model is  
unconnected to that of the hole  argument rekindled by Earman
and Norton \cite{EarNor} and summarized by Earman \cite{Earman2}. (A survey of extensive philosophical debates 
related to this argument is provided by Norton \cite{nortonSEP} and Pooley \cite{pooley}.)

\section{Example of Indeterminism in Cosmology}

We demonstrate that considerations analogous to Norton's Dome arise  naturally through Friedmann's cosmological equations. Our example shows that indeterminism (in the
sense of non-uniqueness of solutions for given initial conditions) 
arises even if nonzero initial conditions are set away from a singularity:
solutions are unique only to the point of failure of the Lipschitz
condition, and can be extended in multiple ways after it. In that sense,
cosmological models extended to include  singularities can have an indeterministic evolution.

With that said, we make no claims regarding the ``physicality" of our
cosmological dome, although the fluid equation of state we use corresponds to the so-called dark energy models 
\cite{steinhardt, darkE}; we view it  as a potentially interesting and attractively simple case of indeterminism arising in the context of classical general relativity.

For deriving the cosmological equations in the form we use, let us note first that in formulating a physical theory, it is common to postulate an action principle for the relevant fields. For general relativity with matter the action is 
$$
S[g,\phi] = \frac{1}{8\pi G}\int d^4x \sqrt{-g} \ R(g) - \int d^4x \sqrt{-g}\ L_M(\phi,g) .
$$
Varying the metric $g$ and the field $\phi$ leads to the general form 
$$
\delta S[g,\phi] = \int d^4x \left( \sqrt{-g}\left( G_{ab} - 8\pi T_{ab}\right)\delta g_{ab} + \sqrt{-g}\  \frac{\delta L_M}{\delta \phi} \delta\phi\right).
$$
The equations of motion derived from the variational principle $\delta S =0$ are the coefficients of $\delta g_{ab}$ and $\delta \phi$. The first of these gives the densitized Einstein equation
$$
\sqrt{-g}\, G_{ab} = 8\pi G \sqrt{-g}\, T_{ab}.
$$
{\it Thus it is this densitized Einstein equation that arises naturally from the variational principle.} If it is assumed that $\sqrt{-g}\ne 0$ everywhere on the manifold, then the more familiar equation $G_{ab} = 8\pi T_{ab}$ follows. 

The solutions spaces of the densitized and undensitized Einstein equations coincide wherever $\det g \ne 0$. But the densitized equations may be well-defined even if $\det g=0$. This is the case for cosmological solutions, a fact we utilize below. (The densitized Einstein equations has been discussed in the literature with application to cosmological solutions \cite{stoica}, and implicitly for quantum fields on  extended FLRW spacetimes \cite{boyle}.)

For a flat universe with metric 
$$ds^2=-dt^2+a^2(t)(dx^2+dy^2+dz^2),$$
$t,x,y,z\in \mathbb{R}$, $a: \mathbb{R}\rightarrow \mathbb{R}$, and perfect fluid source with equation of state $p(t)=w\rho(t)$, the undensitized Einstein equation  leads to  the Friedmann equations 
 \begin{eqnarray}
\frac{\dot{a}^2}{a^2}&=& 2C\rho  \\
\frac{\ddot{a}}{a} &=& C(\beta+2) \rho,
\end{eqnarray}
with $C=4\pi G/3$ and $\beta\equiv-3(1+w)$. These equations are not defined at $a=0$ and so do not permit specification of initial conditions such as $a(0)=0$ and $\dot{a}(0)=0$ at the curvature singularity $a=0$. 

This circumstance is remedied by densitized equations where, for the flat universe metric displayed above, $\sqrt{-g}=a^3(t)$, and the equations are  

\begin{eqnarray}
a\dot{a}^2&=& 2C\tilde{\rho} , \label{rho}\\
a^2 \ddot{a} &=& C(\beta+2)\tilde{\rho},  
\end{eqnarray}
where $\tilde{\rho} \equiv a^3\rho$. Equivalently, we have
\begin{eqnarray}
a^2\ddot{a} - \left( 1+ \frac{\beta}{2}\right) a\dot{a}^2=0,
\label{cosmodome}
\end{eqnarray}
and its solution determines $\tilde{\rho}$ via eqn. (\ref{rho}).   

We now show that  (\ref{cosmodome}) has indeterminate solutions for a certain range of equation of state parameter. First note that,
in addition to the solution $a(t)=0$ for all $t$, 
for $\beta \in (-1,0)$,  (\ref{cosmodome}) has $C^2$ solutions of the 
form 
\begin{equation}
  a(t) = \alpha\lvert t-\kappa\rvert^{-2/\beta} \label{a-soln}
\end{equation}
for all $\alpha$, $\kappa\in \mathbb{R}$. We then have $\tilde{\rho}\sim a^{\beta+3}$, which for $a\neq 0$ agrees with 
the familiar $\rho\sim a^{\beta}=a^{-3(1+w)}$; the density is divergent 
at $t=\kappa$, which is the point of the Big Bang.

Functions (\ref{a-soln}) can be combined to obtain, for any $\kappa\geq 0$, solutions for the same initial conditions $a(0)=\dot{a}(0)=0$:
\begin{equation}
a_\kappa(t) =  \left\{
        \begin{array}{ll}
            0, & t<\kappa  \\
            \alpha(t-\kappa)^{-2/\beta}, & t \geq \kappa
        \end{array}
    \right.
    \label{cos-dome}
\end{equation}
These are the cosmological analogs of Norton's dome.
(In particular, taking $\beta=-\frac{1}{2}$, that is, with the equation
of state parameter $w=-\frac{5}{6}$, the solutions 
correspond mathematically to those in Norton's example (\ref{d-sol}).) Such universes, apparently, can spontaneously enter an acceleration phase from the Big Bang singularity at any time $\kappa \ge 0$.

Similar arguments give examples of multiple solutions, for any
$\kappa\geq 1$, with
nonzero initial conditions at $t=0$: 
\begin{equation}
\label{soln3}
a_{1,\kappa}(t)=\left\{\begin{array}{ll}
\alpha(1-t)^{-2/\beta} & t<1 \\
0 & 1\leq t\leq\kappa \\ 
\alpha(t-\kappa)^{-2/\beta} & t >\kappa 
\end{array}\right.  
\end{equation}
are $C^2$ solutions of (\ref{cosmodome}) with nonzero initial conditions 
$a(0)=\alpha $, $\dot{a}(0)=-\frac{2\alpha}{\beta}$.
Such a universe evolves toward a Big Crunch at $t=1 $, and thereafter 
remains ``dormant" (with a degenerate metric $a=0$) for an arbitrary time 
until a Big Bang at $t=\kappa $.

Mathematically, the indeterminacy arises from the familiar
phenomenon that can be observed in a simple equation such as
$\dot{y}=\sqrt{\lvert y\rvert}$, $y(0)=y_0$: solutions are unique up to the point of failure
of the Lipschitz condition but are not globally unique, with
multiple evolutions possible after that point. 

The range $\beta\in (-1,0)$ gives equation of state parameter $w\in (-1,-2/3)$, a range that 
corresponds to quintessence and other kinds of dark energy \cite{steinhardt}, but excludes the cosmological constant $w=-1$, and  ``phantom energy" defined by $w<-1$. Such equations  of state are used in cosmology to explain the observed late time expansion \cite{darkE}.

\section{Discussion}

The mathematical models presented here raise interesting  philosophical questions. To put it informally, what might one mean by stating that a universe remains ``dormant" over 
a period of coordinate time $t$ with a degenerate metric $a=0$? 
An old paper by  Shoemaker
(\cite{shoemaker}), which discussess the possibility of 
an aparently absurd ``time without change", comes
to mind. Hacking \cite{hacking} gives a conveniently compact
description: 

\begin{quote}
[Shoemaker] takes three shut-off worlds that can observe each other. A and B see C freeze solid, with no change, every $k$ years; A and C see B do it every $m$ years; B and C see A do it every $n$ years. No factors are common to $k$, $m$, and $n$. These propositions are regarded as well corroborated, except that every $kmn$ years no freeze is observed. The simplest explanation is that the whole universe has frozen, and nothing happened while a year elapsed. 
\end{quote}

Our universe during the period $a_{1,\kappa}=0$ (\ref{soln3}) would correspond to the year $kmn$ in Shoemaker's example: 
the interval from $t=1$ to $t=\kappa $ would be a period of 
``coordinate time without change". 

One way to think about this  is from the relational perspective, where the evolution of one dynamical variable is examined relative to another. Examples of such trajectories are the configuration-velocity phase space diagrams of dynamical systems. Variables used in cosmology include matter density, universe volume $V=\lvert a\rvert^3$, and Hubble time defined by $t_H = a/\dot{a}$, with the understanding
that $t_H$ takes the limit value $t_H=0$ where $a=\dot{a}=0$. In terms of the these  relational variables the solution (\ref{soln3}) becomes
\begin{equation}
V(t_H) = \alpha^3 \left\lvert \frac{2}{\beta} t_H\right\rvert^{-6/\beta}.
\label{Vth}
\end{equation}
For example, $\alpha=1$ and $\beta=-2/3$ gives 
$V(t_H)=\lvert 3t_H\rvert^9$; the trajectory is shown in the figure. The passage from $t$ to $t_H$ maps the $t$ interval $[1,\kappa]$ in (\ref{soln3}) to $t_H=0$, hence this mapping is not a bijection, and is not differentiable at $t=1$ and $t=\kappa$; therefore it is not a time reparametrization. And since Hubble time does not tick while the metric is degenerate, {\it the relational view remedies the indeterminacy of equations in coordinate time}. 

There is no such possibility in Newtonian systems with the assumption of 
an external absolute space and time. However, in terms of 
a ``dome time" defined by $t_D=r/\dot{r}$ such that $t_D(r=0)=0$, 
Norton's dome too loses its indeterminacy, as may be seen by rewriting 
$r(t)=r(t(t_D))$ in (\ref{d-sol}). But this would mean that time does not 
tick until the particle moves, a feature perhaps undesirable in Newtonian 
systems. (Nevertheless, there is evidence of an internal gravitational 
time in Newtonian systems of many particles,   a feature that lay 
undiscovered for a few hundred years \cite{Barbour:2014bga}.)

Functions such as $V(t_H)$  are diffeomorphism invariant relational 
observables.  Indeed it is  readily verified that the function $\tilde{V}(t_H)$ is form invariant under the time diffeomorphism $t\longrightarrow f(t)$. The figure also serves to illustrate the perspective that vanishing of volume due to metric degeneracy is not necessarily a manifold pathology.

\begin{figure}[ht]
\begin{center}
   \includegraphics[width=5in]{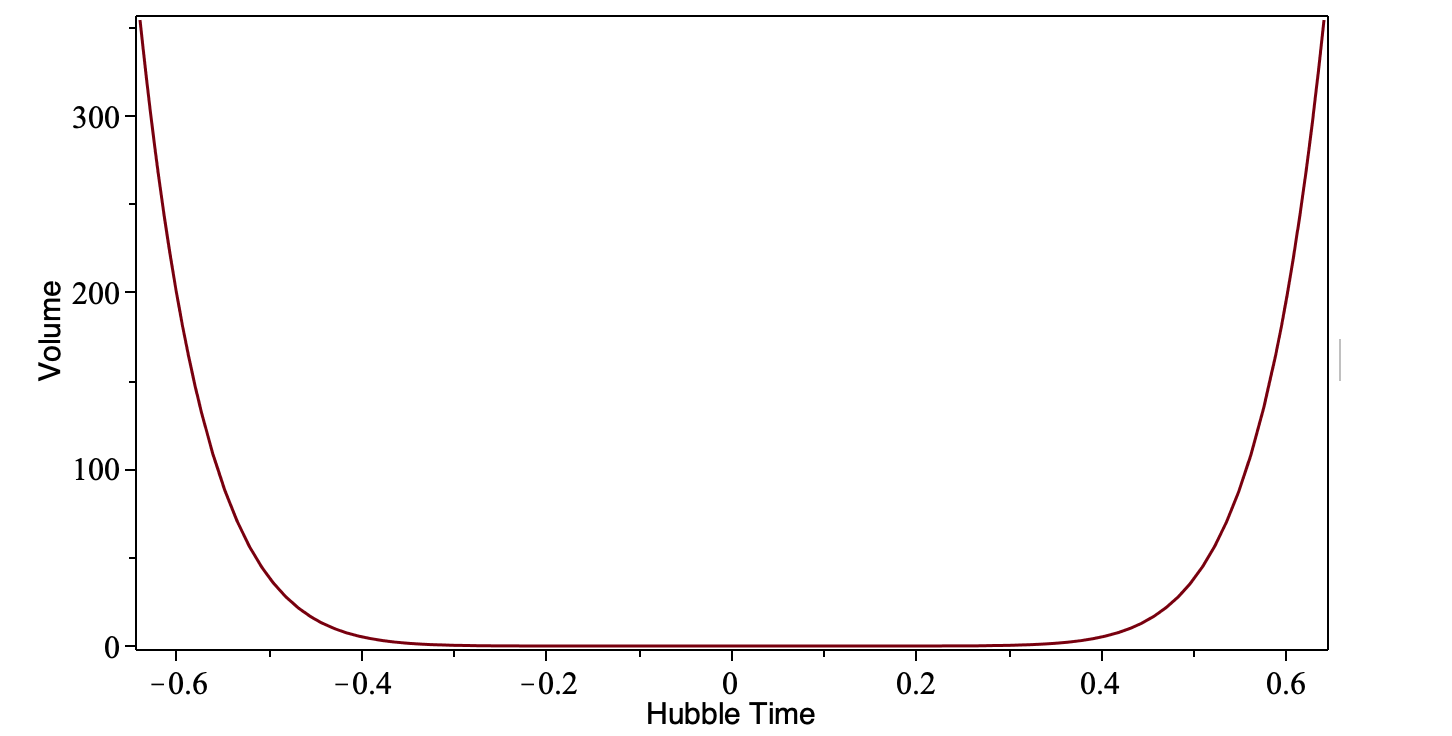}
  \caption{Relational graph of universe volume $\vert a\rvert^3$ as a function of Hubble time $t_H = a/\dot{a}$ for the solution $a=a_{1,\kappa}$ in eqn. (\ref{soln3}) and $V(t_H)$ in eqn. (\ref{Vth}) with $\alpha=1$ and $\beta=-2/3$; the figure is analogous to standard configuration vs. velocity phase space plots of dynamical systems.}
\end{center}
\end{figure}

In comparing Norton's dome with our cosmological example, we note the dome's intrinsic metric and Gaussian curvature $K(r)$ (from the parametrization given above): 
\[
ds^2 = dr^2 + R(r)^2 d\phi^2 ; \quad  K(r) =  \frac{4}{3}\left(\sqrt{1-r} - (1-r)^2\right)^{-1}. 
\]
Since $\lim_{r\rightarrow 0^+}R(r)=0$, the metric is degenerate at the 
dome's apex, and the fact that $\lim_{r\rightarrow 0^+}K(r)=\infty$ 
establishes that this point is a singularity of the Gaussian curvature. 
This is analogous to the divergence of the Riemann curvature at $a=0$.
Furthermore, the arc length traversed by a particle along the dome is a physical observable evolving with respect to Newtonian absolute time, a feature similar to the volume of the universe evolving with respect to Hubble time.

From a  mathematical perspective these  cosmological models exhibit indeterminism, in the form of non-uniqueness of solutions for given initial conditions, of the Einstein field equations over an extended spacetime. The densitized equations, in our view, can be justified using arguments generally accepted in mathematical physics, and extended spacetimes have indeed been considered in literature. From the perspective of the Hamiltonian formulation of general relativity \cite{Wald:1984rg}, the densitized equations may be viewed as coming from a special choice of lapse function.

Our objective here is not to recount the rich  
debate on the ontological status of spacetime, or to enumerate
the effects of various positions in the debate might have on our examples, but to introduce what seem to us some new aspects to consider. We summarize them in the following paragraphs.

We already indicated that the model exhibited here differs from the general considerations arising from the hole argument. Further,  previously studied concrete examples of indeterminism in Newtonian mechanics, 
special relativity, and geodesic motion on 
curved spacetime all arise in the context of a preset spacetime stage, 
whether it is Newton's absolute space 
and time, or an externally provided pseudo-Riemannian metric. Instances of such scenarios arise for massive particle geodesics  on a specially configured spherically symmetric spacetime background,  and for charged particle motion in Minkowski spacetime with a contrived force; in both cases certain test particle trajectories  are given by an equation of the type $\ddot{r} =\sqrt{r}$ \cite{fletcher2}.

In contrast, the cosmological model considered here is of an 
ontologically different nature in that it provides an example where  
a mathematical description of the evolution of spacetime itself is 
indeterminate.

Additionally the model avoids many of the objections that might be raised for a Newtonian system, such as violation of 
the First Law, loss of contact with the dome, or an incomplete 
formulation of Newtonian theory. One may, of course, object to 
``physicality" of 
a mathematical example we present. However, the purpose of our model is not to give an accurate description of the 
observable universe, but rather to highlight that a kind of matter used to model a phase of it 
leads to indeterminacy if the equation of state parameter 
is a constant in the range $w \in (-1,-2/3)$.  

Finally, it should also be noted that the physical universe is expected to be quantum in nature near $a=0$, where GR would not be expected to apply. There are effective semiclassical models of the universe that resolve the singularity and replace it with a ``bounce." These have  solutions where $a(0)= a_0>0$ and $\dot{a}(0)=0$.  Whether such models exhibit indeterminacy of the type we have explored here is an interesting question, for it would make the future evolution non-unique starting from a point of finite spacetime curvature.

\bigskip

\noindent\underbar{Acknowledgements} We thank John Norton for helpful comments on an earlier draft of this paper. This work was supported in part by the Natural Science and Engineering Research Council of Canada.

\bigskip

\bibliography{domeref}

\end{document}